# Limited Feedback for Multi-Antenna Multi-user Communications with Generalized Multi-Unitary Decomposition

Wee Seng Chua[1], Chau Yuen[2], Yong Liang Guan[1], and Francois Chin[2]
[1]Nanyang Technological University, [2]Institute for Infocomm Research, Singapore
Email: chua0159@ntu.edu.sg, cyuen@i2r.a-star.edu.sg

*Abstract* – In this paper, we propose a decomposition method called Generalized Multi-Unitary Decomposition (GMUD) which is useful in multi-user MIMO precoding. This decomposition transforms a complex matrix $\mathbf{H} \in \mathbb{C}^{m \times n}$ into $\mathbf{H} = \mathbf{P}_i \mathbf{R} \mathbf{Q}_i^H$, where $\mathbf{R}$ is a special matrix whose first row contains only a non-zero user defined value at the left-most position, $\mathbf{P}_i$ and $\mathbf{Q}_i$ are a pair of unitary matrices. The major attraction of our proposed GMUD is we can obtain multiple solutions of $\mathbf{P}_i$ and $\mathbf{Q}_i$. With GMUD, we propose a precoding method for a MIMO multi-user system that does not require full channel state information (CSI) at the transmitter. The proposed precoding method uses the multiple unitary matrices property to compensate the inaccurate feedback information as the transmitter can steer the transmission beams of individual users such that the inter-user interference is kept minimum.

## I. INTRODUCTION

Multiple input multiple output (MIMO) schemes have assumed great importance in current and next generation broadband wireless networks. They have been used for increasing spatial diversity, spatial multiplexing, beamforming and precoding.

In a multi-user broadcast communication system, where the base station with multiple antennas simultaneously sends one data stream to each user with one receive antenna, the regularize-inverse precoding proposed in [1] can improve the probability of bit errors, and combine it with vector-perturbation as shown in [2] can further reduce the energy of the transmitted signal so as to achieve excellent sum-rate with low bit errors. In addition, we show that by having additional receive antennas at the user, which is feasible in many future communication systems e.g. 3Gpp LTE, the performance in probability of bit errors can significantly improve if the transmitter can select the optimum receive antenna per user when sending one data stream per user. However, this selection can only be performed if full CSI knowledge is available at the transmitter, and it is a tricky task to feedback full CSI from all users to the transmitter. Thus, it is important to consider both the probability of bit error rate and the amount of feedback information in the design of the precoding matrix. Apparently, there is a tradeoff between the performance and the amount of feedback information, and it is a challenge to maximize the advantage of having additional antenna to improve the performance when the feedback information of the channel information is not complete.

In this paper, we propose a precoding technique that makes use of the additional antenna without increasing the feedback information. This is analogous to what is mentioned in [3], when the increasing number of receive antennas is not used to increase the number of data streams received at each user, it can increase the channel estimation quality.

## II. SIGNAL MODEL

Given $N_T$ transmit antennas at the base station servicing one data stream to each of a pool of $K$ users with $N_R$ receive antennas each, the received signal at $k^{th}$ user is given as

$$\mathbf{y}_k = \mathbf{H}_k \mathbf{x} + \mathbf{n}_k \quad (1)$$

where $\mathbf{x}$ represents the transmit signal vector, $\mathbf{H}_k$ is a $N_R \times N_T$ matrix containing the channel information between the base station and $k^{th}$ user, $\mathbf{n}_k$ is the additive white Gaussian noise vector such that $E\mathbf{n}_k \mathbf{n}_k^H = \sigma^2 \mathbf{I}$. The transmitter signal $\mathbf{x}$ from the transmitter can be represented as

$$\mathbf{x} = (\mathbf{G}\mathbf{u})/\sqrt{\gamma} \quad (2)$$

where $\mathbf{G}$ is a $N_T \times K$ precoding matrix, $\mathbf{u} = [u_1 \cdots u_K]^T$ whose element $u_k$ is the intended data for $k^{th}$ user, and $\gamma = \|\mathbf{G}\mathbf{u}\|^2$ is used to normalize the transmitted signal.

In a multi-user with one receive antenna communication, the regularize-inverse precoding in [1] arranges the channel information into a $K \times N_T$ matrix such that each individual row of $\tilde{\mathbf{H}}$ represents its corresponding channel information between the base station and its respective user. The precoding matrix $\mathbf{G}$ is

$$\mathbf{G}_{\text{reg-inv}} = \tilde{\mathbf{H}}^H \left( \tilde{\mathbf{H}} \tilde{\mathbf{H}}^H + \alpha \mathbf{I} \right)^{-1} \quad (3)$$

where $\alpha = K/\rho = K\sigma^2$ as derived in [1], $\mathbf{I}$ is a $K \times K$ identity matrix and $[.]^H$ denotes complex conjugate.

Although this precoding matrix introduces inter-user interference due to $\tilde{\mathbf{H}} \tilde{\mathbf{H}}^H \left( \tilde{\mathbf{H}} \tilde{\mathbf{H}}^H + \alpha \mathbf{I} \right)^{-1} \neq \mathbf{I}$, it reduces the normalization constant, $\gamma$, significantly, which results in a stronger signal of higher SINR at the receiver. The amount of interference is determined by $\alpha$.

When more than one receive antenna is used to receive one data stream, the communication channel can be arranged into $K \times N_T$ matrix, whose rows represent the corresponding channel information between the $N_T$ transmit antennas and one of the $N_R$ receive antennas of its respective user, and there are a total of $(N_R)^K$ different combinations. The regularize-inverse precoding matrix is thus becoming

$$\hat{\mathbf{G}}_{\text{reg-inv}} = \hat{\mathbf{H}}^H \left( \hat{\mathbf{H}}\hat{\mathbf{H}}^H + \alpha \mathbf{I} \right)^{-1} \quad (4)$$

where $\hat{\mathbf{H}}$ is one of the $(N_R)^K$ different combinations. The received signal vector $\mathbf{y} = [\mathbf{y}_1 \quad \cdots \quad \mathbf{y}_K]^T$ becomes

$$\mathbf{y} = \frac{\hat{\mathbf{H}}\hat{\mathbf{H}}^H \left( \hat{\mathbf{H}}\hat{\mathbf{H}}^H + \alpha \mathbf{I} \right)^{-1} \mathbf{u}}{\sqrt{\gamma}} + \mathbf{n} = \hat{\mathbf{H}}_{\text{eq}}\mathbf{u} + \mathbf{n} \quad (5)$$

where $\hat{\mathbf{H}}_{\text{eq}} = \dfrac{\hat{\mathbf{H}}\hat{\mathbf{H}}^H \left( \hat{\mathbf{H}}\hat{\mathbf{H}}^H + \alpha \mathbf{I} \right)^{-1}}{\sqrt{\gamma}} = \begin{bmatrix} \hat{h}_{1,1} & \cdots & \hat{h}_{1,K} \\ & \ddots & \\ \hat{h}_{K,1} & & \hat{h}_{K,K} \end{bmatrix}$.

The regularize-inverse precoding with receive antenna selection can be obtained by maximizing min signal to interference plus noise ratio of the received signals.

$$\hat{\mathbf{G}}_{\text{reg-inv,optimal}} = \max_{\hat{\mathbf{H}}} \min_{1 \leq m \leq K} \left( \frac{|\hat{h}_{m,m}|^2}{\sum_{n=1, n \neq m}^{K} |\hat{h}_{m,n}|^2 + \gamma/\rho} \right) \quad (6)$$

## III. GENERALIZED MULTI-UNITARY DECOMPOSITION (GMUD)

Given a complex matrix channel $\mathbf{H}$, it can transform into various forms using different decomposition techniques such as Singular Value Decomposition (SVD) [4], Geometric Mean Decomposition (GMD) [5,6] and Geometric Triangular Decomposition (GTD) [7].

In this paper, we propose a new decomposition technique called *Generalized Multi-Unitary Decomposition (GMUD)*, that transforms $\mathbf{H}$ to $\mathbf{P}_i \mathbf{R} \mathbf{Q}_i^H$, where $\mathbf{R}$ can be a $R \times R$ lower triangular matrix or a special $R \times R$ matrix with a prescribed value at the first element and zeros for the rest of the elements in the first row, $\mathbf{P}_i$ and $\mathbf{Q}_i$ are a group of different unitary matrices. It is a general decomposition method that includes SVD, GMD, and GTD as part of the solutions. Both GMUD and GTD allow the user to prescribe the diagonal values of the $\mathbf{R}$ matrices. However, GMUD can produce more than one pairs of different unitary matrices, as opposed to one pair produced by GTD and the other decompositions.

### A. Derivation of R

Consider a rank $R$ complex $\mathbf{H}$ with singular values $\lambda_R \leq r \leq \lambda_1$. $\mathbf{R}$ can be defined as a special $R \times R$ matrix in the form of

$$\mathbf{R} = \begin{bmatrix} r & 0 & \cdots & 0 \\ c_{21} & c_{22} & \cdots & c_{2R} \\ \vdots & \vdots & \vdots & \vdots \\ c_{R1} & c_{R2} & \cdots & c_{RR} \end{bmatrix} \quad \text{where } \lambda_R \leq r \leq \lambda_1 \quad (7)$$

In the first row of $\mathbf{R}$, there is only one non-zero positive element $r$ at the (1,1) position. The remaining elements at the other rows are calculated based on $r$ and the singular values. $\mathbf{R}$ can also be defined as a lower triangular matrix with user pre-defined diagonal elements. This can be achieved by assigning the next remaining row in the same way after the previous row has been assigned and let all the entries on the right of the diagonal entries to be zero. Thus this leads to

$$\mathbf{R} = \begin{bmatrix} r_1 & 0 & \cdots & 0 \\ c_{21} & r_2 & 0 & 0 \\ \vdots & \vdots & \vdots & \vdots \\ c_{R1} & c_{R2} & \cdots & r_R \end{bmatrix} \quad (8)$$

where the values of the diagonal values follow the condition as described in [7].

From this point onwards, for simplicity of illustration, we consider $R = 2$, however GMUD can be extended to higher rank. The channel $\mathbf{H}$ is first transforms into $\mathbf{H} = \mathbf{U}\mathbf{\Lambda}\mathbf{V}^H$ using SVD. The matrix $\mathbf{R}$ with a pre-assigned value $\lambda_R \leq r \leq \lambda_1$ at the (1,1) position can be arranged in the form of $\mathbf{R} = \mathbf{U}_0^H \mathbf{\Sigma} \mathbf{V}_0$, where $\mathbf{U}_0$ and $\mathbf{V}_0$ are unitary matrices assigned using Givens rotations, $\mathbf{\Lambda}$ is the same diagonal matrix of $\mathbf{H}$ containing the singular values.

$$\mathbf{U}_0 = \begin{bmatrix} a^H & b^H \\ -b^H & a^H \end{bmatrix}, \mathbf{V}_0 = \begin{bmatrix} c & s \\ -s & c \end{bmatrix} \quad (9)$$

where $b = \sqrt{1-a^2}$, and $s = \sqrt{1-c^2}$. As a result, $r$, $z_1$ and $z_2$ can be defined in terms of $a$, $b$, $c$, $s$, $\lambda_1$ and $\lambda_2$.

$$\mathbf{R} = \mathbf{U}_0^H \mathbf{\Lambda} \mathbf{V}_0$$

$$\begin{bmatrix} r & 0 \\ z_1 & z_2 \end{bmatrix} = \begin{bmatrix} a^H & b^H \\ -b^H & a^H \end{bmatrix}^H \begin{bmatrix} \lambda_1 & 0 \\ 0 & \lambda_2 \end{bmatrix} \begin{bmatrix} c & s \\ -s & c \end{bmatrix} \quad (10)$$

$$\begin{bmatrix} r & 0 \\ z_1 & z_2 \end{bmatrix} = \begin{bmatrix} ac\lambda_1 + bs\lambda_2 & as\lambda_1 - bc\lambda_2 \\ bc\lambda_1 - as\lambda_2 & bs\lambda_1 + ac\lambda_2 \end{bmatrix}$$

The value of $a$ and $c$ can be derived from the first row of (10), after substituting $b$ and $s$ from (9), and subsequently $\mathbf{U}_0$ and $\mathbf{V}_0$ can be determined.

$$ac\lambda_1 + bs\lambda_2 = r, \quad as\lambda_1 - bc\lambda_2 = 0 \quad (11)$$

The remaining values of $\mathbf{R}$ can be found by substituting the values of $a$ and $c$

$$z_1 = bc\lambda_1 - as\lambda_2, \quad z_2 = bs\lambda_1 + ac\lambda_2 \quad (12)$$

Next, replacing $\mathbf{\Lambda} = \mathbf{U}_0 \mathbf{R} \mathbf{V}_0^H$ into the singular value decomposition of $\mathbf{H}$ yields

$$\mathbf{H} = (\mathbf{U}\mathbf{U}_0) \mathbf{R} (\mathbf{V}\mathbf{V}_0)^H = \mathbf{P}_0 \mathbf{R} \mathbf{Q}_0^H \quad (13)$$

Since $\mathbf{U}, \mathbf{U}_0, \mathbf{V}$ and $\mathbf{V}_0$ are unitary matrices, $\mathbf{P}_0$ and $\mathbf{Q}_0$ will also be unitary matrices.

## B. Derivation of $P_i$ and $Q_i$

In order to get multiple different **P** and **Q**, we include $\mathbf{M}_i$, a diagonal matrix with some unity phase at the diagonal elements, and its conjugate transpose $\mathbf{M}_i^H$ to (13), where $\mathbf{M}_i \mathbf{\Lambda} \mathbf{M}_i^H = \mathbf{\Lambda}$.

$$\mathbf{M}_i = \begin{bmatrix} e^{j\theta_{i1}} & 0 \\ 0 & e^{j\theta_{i2}} \end{bmatrix}, \quad \mathbf{M}_i^H = \begin{bmatrix} e^{-j\theta_{i1}} & 0 \\ 0 & e^{-j\theta_{i2}} \end{bmatrix} \quad (14)$$

where $\theta_{i1}$ and $\theta_{i2}$ can be any value from 0 to $2\pi$. After the inclusion of $\mathbf{M}_i$ and $\mathbf{M}_i^H$, **H** becomes

$$\mathbf{H} = (\mathbf{U}\mathbf{M}_i \mathbf{U}_0)\mathbf{R}(\mathbf{V}\mathbf{M}_i \mathbf{V}_0)^H = \mathbf{P}_i \mathbf{R} \mathbf{Q}_i^H \quad (15)$$

From (15), **R** is independent to the value of $\mathbf{M}_i$, hence it remains the same. Since $\mathbf{P}_i$ and $\mathbf{Q}_i$ include the information of $\mathbf{M}_i$, it is apparent that $\mathbf{P}_i$ and $\mathbf{Q}_i$ are varying with $\theta_{i1}$ and $\theta_{i2}$. In addition, $\mathbf{U}, \mathbf{U}_0, \mathbf{V}$ and $\mathbf{V}_0$ are unitary matrices, the combination of other unitary matrices $\mathbf{M}_i$ and $\mathbf{M}_i^H$, will result in $\mathbf{P}_i$ and $\mathbf{Q}_i$ being unitary matrices as well.

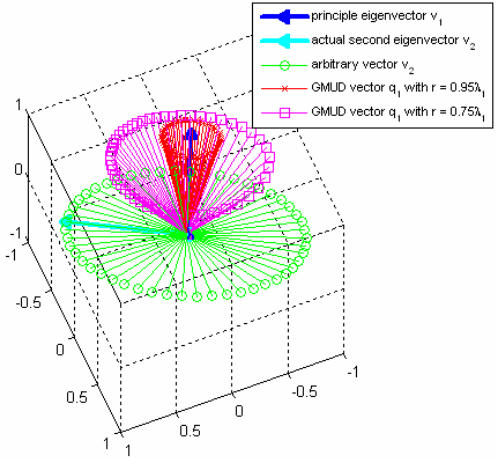

Figure 1: The principle and second eigen-vectors of SVD and the first column vectors of the unitary matrix $\mathbf{Q}_i$ of GMUD with different $\mathbf{M}_i$ are plotted on a uniform 3-D sphere. It shows that with the same **R**, different $\theta_{i1}$ or $\theta_{i2}$ produces different vectors in $\mathbf{Q}_i$, which form a cone whose center is the principle eigen-vector of SVD.

For the case of dimension of two, we only need to know the principle eigenvector, i.e. the first row of $\mathbf{V}_0$. The exact second eigenvector is not important, as eventually the rotation matrix $\mathbf{M}_i$ will make a rotation and generate a series of vectors. Hence the decomposition of a 2-by-2 matrix only can be performed with any vector that is orthogonal to the principal eigenvector, not necessary with the exact second eigenvector. So we can generate an arbitrary second eigenvector that is orthogonal to the principal eigenvector, as shown in Figure 1.

An example is shown in Figure 1, different $\theta_{i1}$ with values ranging from 0 to $2\pi$, produce different $\mathbf{M}_i$, which result in different vectors. These vectors form a cone surrounding the principle eigen-vector of SVD which is located at the center of the cone. The value of $r$ determines the radius of the cone, i.e. the closer the value of $r$ to the largest singular value $\lambda_1$, the smaller the cone radius. If the value of $r$ is equal to $\lambda_1$, the first column vector of $\mathbf{Q}_i$ becomes the principle eigenvector of SVD. The lines with square markers represent GMUD vectors with a smaller $r = 0.75 \lambda_1$, while the lines with asterisk markers represent GMUD vectors with a larger $r = 0.95 \lambda_1$.

## IV. PRECODING BASED ON GMUD

In the scenario where there are $N_T$ transmit antennas at the base station sending one data stream to each of the $K$ users with $N_R$ receive antennas each. The received signal per user is given in (1) and its corresponding channel matrix $\mathbf{H}_k$ is decomposed using GMUD, multiple different pairs of $\mathbf{P}_i$ and $\mathbf{Q}_i$ matrices can be generated with the same **R** matrix, and each user's $\mathbf{H}_k$ can be decomposed into

$$\mathbf{H}_k = \mathbf{P}_k \mathbf{R}_k \mathbf{Q}_k^H \quad (16)$$

where we replace $r$ in **R** from (7) and $\theta_{i1}$ in $\mathbf{M}_i$ from (14) (for the case of 2x2 matrices, the optimization of single parameter $\theta_{i1}$ is enough) with optimizing parameters $r_k$ and $\theta_k$ respectively, and we denote the optimized $\mathbf{P}_i$, $\mathbf{Q}_i$ and **R** as $\mathbf{P}_k$, $\mathbf{Q}_k$ and $\mathbf{R}_k$ respectively. The first column vector of $\mathbf{Q}_{r_k,\theta_k}$ is considered as an individual transmission beam for that user. In other words, it is important to have many different $\mathbf{Q}_k$ matrices, which represent different transmission beams, because the transmitter can steer the beams of every users to make them as orthogonal as possible. If all the users' transmission beams are orthogonal to each other, there will be zero multi-users interference. Hence by changing $r_k$ and $\theta_k$, multiple beamforming vectors "listening" to different direction can be obtained.

The precoding matrix **G** can be formed by assigning each column of the matrix as the first column vectors of $\mathbf{Q}_k$ of each users. From (15) and (16),

$$\mathbf{Q}_k = \mathbf{V}\mathbf{M}_{\theta_k}\mathbf{V}_0 = \begin{bmatrix} \underbrace{\mathbf{q}_{1,k}}_{\mathbf{g}_k} & \mathbf{q}_{2,k} \end{bmatrix} \quad (17)$$

where $\mathbf{q}_{1,k}$ and $\mathbf{q}_{2,k}$ are the first and second column vectors respectively.

For simplicity, we shall focus on the example of having $K = N_T = N_R = 2$ for the rest of the document. The idea is extendable to any number of transmit antennas or any users with any number of receive antennas. We let

$$\mathbf{g}_k = \mathbf{q}_{1,k}, \quad \mathbf{g}_l = \mathbf{q}_{1,l} \quad (18)$$

and the precoding matrix is

$$\mathbf{G} = [\mathbf{g}_k \quad \mathbf{g}_l] \quad (19)$$

When we combine (1), (16) and (19), the received signal for user $k$ becomes

$$\mathbf{y}_k = \mathbf{P}_k \mathbf{R}_k \mathbf{Q}_k^H \frac{\mathbf{Gu}}{\sqrt{\gamma}} + \mathbf{n}_k = \mathbf{P}_k \mathbf{R}_k \begin{bmatrix} \mathbf{q}_{1,k}^H \\ \mathbf{q}_{2,k}^H \end{bmatrix} \begin{bmatrix} \mathbf{g}_k & \mathbf{g}_l \end{bmatrix} \frac{\mathbf{u}}{\sqrt{\gamma}} + \mathbf{n}_k$$

$$= \frac{1}{\sqrt{\gamma}} \mathbf{P}_k \mathbf{R}_k \begin{bmatrix} \mathbf{q}_{1,k}^H \mathbf{g}_k u_k + \mathbf{q}_{1,k}^H \mathbf{g}_l u_l \\ \mathbf{q}_{2,k}^H \mathbf{g}_k u_k + \mathbf{q}_{2,k}^H \mathbf{g}_l u_l \end{bmatrix} + \mathbf{n}_k$$

(20)

where $\gamma = \|\mathbf{Gu}\|^2$ is used to normalized the transmitted signal.

Given $\mathbf{R}_k$ is a special matrix as shown in (7) and (8), (20) can be reduced to

$$\mathbf{y}_k = \frac{1}{\sqrt{\gamma}} \mathbf{P}_k \begin{bmatrix} r_k \left( \mathbf{q}_{1,k}^H \mathbf{g}_k u_k + \mathbf{q}_{1,k}^H \mathbf{g}_l u_l \right) \\ \varepsilon \end{bmatrix} + \mathbf{n}_k \quad (21)$$

where $\varepsilon = c_{21,k} \left( \mathbf{q}_{1,k}^H \mathbf{g}_k u_k + \mathbf{q}_{1,k}^H \mathbf{g}_l u_l \right) + c_{22,k} \left( \mathbf{q}_{2,k}^H \mathbf{g}_k u_k + \mathbf{q}_{2,k}^H \mathbf{g}_l u_l \right)$, and $c_{21,k}$ and $c_{22,k}$ refer to the elements of $\mathbf{R}_k$ in (7).

In order to reduce the BER, the interference terms must be minimized by choosing $\mathbf{q}_{1,k}$ and $\mathbf{g}_l$ to be as orthogonal to each other as possible. However, it is more appropriate to use the cost function of maximizing the signal to interference-plus-noise ratio (SINR) to find the precoding matrix $\mathbf{G}$ where the signal is $r_k u_k / \sqrt{\gamma}$ and the interference is $r_k \mathbf{q}_{1,k}^H \mathbf{g}_l u_l / \sqrt{\gamma}$. This is because reducing the dot product between two users' $\mathbf{q}_1$ vectors may produces a $\mathbf{G}$ that has a large normalization constant $\gamma$ which results in a weaker received signal and increases the probability of bit error at the receiver. Thus, the cost function of finding $\mathbf{G}$ becomes

$$\mathbf{G} = \max_{\mathbf{g}_k, \mathbf{g}_l} \min \left( \frac{|r_k|^2}{|r_k|^2 |\mathbf{q}_{1,k}^H \mathbf{g}_l|^2 + \sigma^2 \gamma}, \frac{|r_l|^2}{|r_l|^2 |\mathbf{q}_{1,l}^H \mathbf{g}_k|^2 + \sigma^2 \gamma} \right)$$

$$= \max_{r_k, r_l, \theta_k, \theta_l} \min \left( \frac{|r_k|^2}{|r_k|^2 |\mathbf{q}_{1,k}^H \mathbf{q}_{1,l}|^2 + \sigma^2 \gamma}, \frac{|r_l|^2}{|r_l|^2 |\mathbf{q}_{1,l}^H \mathbf{q}_{1,k}|^2 + \sigma^2 \gamma} \right)$$

(22)

The performance of the system improves if we choose different power loading factor to $\mathbf{g}_k$ and $\mathbf{g}_l$ and $\mathbf{G}$ becomes

$$\mathbf{G} = \begin{bmatrix} \alpha \mathbf{g}_k & \beta \mathbf{g}_l \end{bmatrix} \quad (23)$$

Thus the cost function becomes

$$\mathbf{G} = \max_{r_k, r_l, \theta_k, \theta_l, a, b} \min \left( \frac{\alpha^2 |r_k|^2}{\beta^2 |r_k|^2 |\mathbf{q}_{1,k}^H \mathbf{q}_{1,l}|^2 + \sigma^2 \gamma}, \frac{\beta^2 |r_l|^2}{\alpha^2 |r_l|^2 |\mathbf{q}_{1,l}^H \mathbf{q}_{1,k}|^2 + \sigma^2 \gamma} \right)$$

(24)

Since we assume the receiver only feedback the principal eigenvector, in (22) and (24) the precoding matrix generated by the transmitter does not make use of the second column vector of $\mathbf{Q}_i$. In other words, the receiver decodes the received signal without the use of $\varepsilon$ as shown in (21).

Hence the proposed GMUD precoding only requires *partial CSI feedback,* i.e. it only requires the receiver to feedback the eigenvalues and eigenvectors. For the case of 2x2, the feedback information can be further reduced to the singular values and principal eigenvector.

## V. SIMULATION RESULTS AND DISCUSSION

The performance of the MIMO multi-user system using GMUD precoding and regularize-inverse precoding [1] with or without antenna selection is compared under perfect and quantized limited feedback information in this section. Given $N_T$ transmit antennas at the base station sending one data stream to each of the $K$ users with $N_R$ receive antennas each. For the regularize-inverse precoding with antenna selection at the transmitter, the full channel matrix of the $k^{th}$ user, $\mathbf{H}_k$, in (1) is needed to be feedback from all users to the transmitter, this can be done by normalizing its row or column vector first, and feedback the normalizing scalars and unit norm vectors. For regularize-inverse precoding without antenna selection, the user needs to feedback one row of $\mathbf{H}$ to the transmitter, likewise this can be done by normalizing this row vector and feedback the normalizing scalar and the unit norm vector. For the proposed precoding based on GMUD, the user will send back the singular values and the first column vector of $\mathbf{V}$. The number of equivalent scalars needed to be feedback per user for the above three different schemes are summarized in Table 1 for the case of a MIMO system with two transmit antennas at the transmitter, and two receive antennas per user.

In the simulation, we consider a system with two transmit antennas, and two users with two receive antennas each. We are using the same total number of feedback bits (besides the unquantized case) for each scheme to test their vulnerability to imperfect feedback channel information. In the case of regularize-inverse precoding with antenna selection, we are using $N$ bits and $2N$ bits to quantize each of the real normalized element and normalizing constant respectively. If there is no antenna selection, we are using double number of bits for all the information compared to the antenna selection case. For GMUD precoding, we are allocating $2N$ bits for all the information. We are using a total of $12N$ bits to quantize the feedback information. For example, if we are using 48 feedback bits, $N$ will become 4.

**Table 1: Allocation of the bits used to quantize the feedback information**

| Feedback Info / Precoding Scheme | No. of equivalent scalars for unit norm vector (Bits used) | No. of normalizing scalar or singular values (Bits used) |
|---|---|---|
| Reg-Inv with antenna selection | 8 ($N$ bits) | 2 ($2N$ bits) |
| Reg-Inv without antenna selection | 4 ($2N$ bits) | 1 ($4N$ bits) |
| GMUD | 4 ($2N$ bits) | 2 ($2N$ bits) |

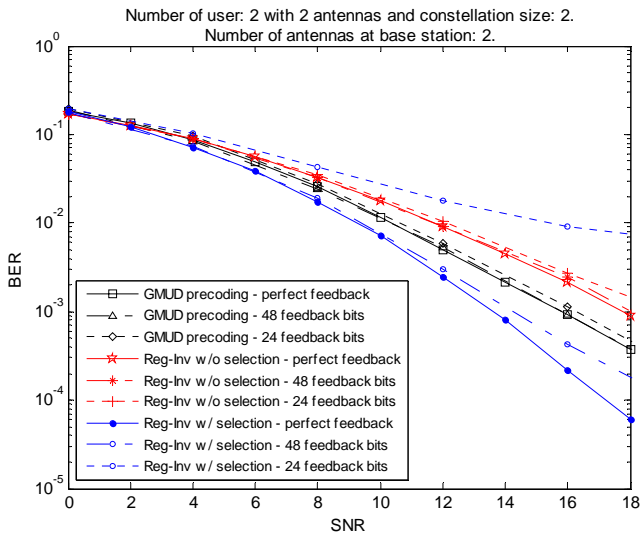

**Figure 2: Performance of probability of bit error for 2 users using QPSK symbols given perfect or quantized feedback information (48, 24 bits) available at the transmitter.**

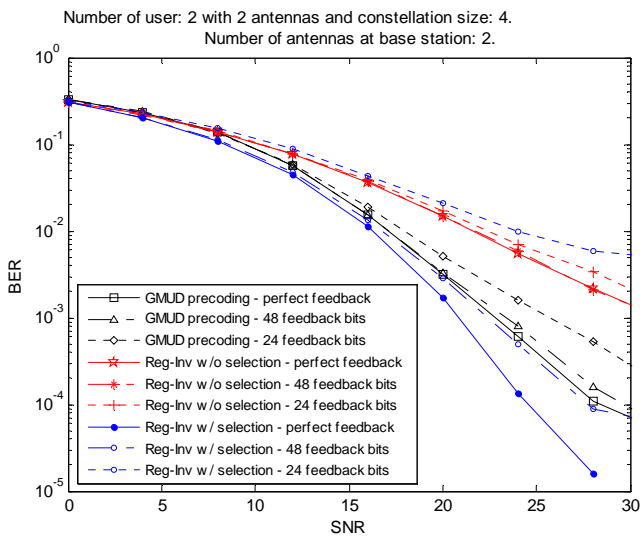

**Figure 3: Performance of probability of bit error for 2 users using 16QAM symbols given perfect or quantized feedback information (48, 24 bits) available at the transmitter.**

Figure 2 and Figure 3 show that under the assumption of having perfect CSI at the transmitter, regularize-inverse precoding with antenna selection requires double the feedback information than regularize-inverse precoding without antenna system to obtain a gain of around 4.1dB and 10.75dB respectively, while GMUD precoding uses nearly half feedback information to achieve a gain of around 1.95dB and 8.8dB respectively. When the three schemes have the same number of feedback bits as shown in Table 1, the BER performance of regularize-inverse precoding with antenna selection deteriorates tremendously if limited feedback information is available at the transmitter. As shown in Figure 3, when there are only 48 bits available, the performance of regularize-inverse precoding suffers a 1.75dB loss as compare to perfect CSI, the lost widen as the feedback information becomes more scarce, and eventually an error floor can be seen at BER of 0.05. However, GMUD precoding does not depend too much on the accuracy of the feedback, it suffers a negligible loss of when limited feedback is available, thus making it a robust scheme for limited feedback communication.

We would like to mention that the number of bits used in this paper is just for illustration purposes. In practice, one can use other ways to feedback the channel matrix, e.g. codebook based or compression based on Givens rotation. Our intention is to show that with limited feedback constraint, it is important to feedback less information but with better quality, rather than feedback a lot of information but with poor quality.

## VI. CONCLUSION

GMUD precoding provides an alternative to feedback less information through mathematical decomposition method, while maintaining a high level of BER performance. It can decompose a $m \times n$ channel matrix into multiple $m \times m$ and $n \times n$ unitary matrices, and a $m \times n$ triangular matrix with prescribed diagonal elements. Applying the multi-unitary matrices property of GMUD, the transmitter can steer the transmission beams of individual users such that the inter-user interference is kept minimum. We show that the performance of GMUD does not deteriorate when the feedback information becomes limited.